\documentclass[useAMS,usenatbib,usegraphicx]{mn2e}
\usepackage{lscape}
\usepackage{multirow}
\usepackage{natbib}
\usepackage{graphicx}
\usepackage{float}
\usepackage{color}

\newcommand{\ee}[1]{\mbox{${} \times 10^{#1}$}}

\newcommand{\bl}[1]{\mbox{\boldmath$ #1 $}}

\newcommand{\degree}{\mbox{$^{\circ}$}}

\newcommand{\um}{$\mu$m}
\newcommand{\msun}{\mbox{M$_\odot$}}

\newcommand{\mcore}{\mbox{$M_{core}$}}
\newcommand{\mdisc}{\mbox{$M_{disc}$}}
\newcommand{\mstar}{\mbox{$M_{star}$}}
\newcommand{\renv}{\mbox{$r_{core}$}}
\newcommand{\rdisc}{\mbox{$R_{disc}$}}
\newcommand{\mdotstar}{\mbox{$\dot{M}_s$}}
\newcommand{\mdotdisc}{\mbox{$\dot{M}_d$}}
\newcommand{\kl}{\mbox{k${\lambda}$}}

\begin{document}
\title{On the reliability of protostellar disc mass measurements and the existence of fragmenting discs}

\author[Dunham et al.]
{
Michael M.~Dunham$^{1,2}$, Eduard I.~Vorobyov$^{3,4}$, \& H\'ector G.~Arce$^{5}$ \\
$^{1}$ Harvard-Smithsonian Center for Astrophysics, 60 Garden Street, MS 78, Cambridge, MA 02138, USA \\
$^{2}$ mdunham@cfa.harvard.edu \\
$^{3}$ Department of Astrophysics, University of Vienna, Vienna, 1180, Austria \\
$^{4}$ Research Institute of Physics, Southern Federal University, Rostov-on-Don, 344090, Russia \\
$^{5}$ Department of Astronomy, Yale University, P.O. Box 208101, New Haven, CT 06520, USA\\
}

\maketitle

\begin{abstract}
We couple non-magnetic, hydrodynamical simulations of collapsing protostellar 
cores with radiative transfer evolutionary models to generate 
synthetic observations.  We then use these synthetic observations to 
investigate the extent to which a simple method for measuring protostellar 
disc masses used in the literature recovers the intrinsic masses of the 
discs formed in the simulations.  We evaluate the effects of contamination 
from the surrounding core, partially resolving out the disc, optical depth, 
fixed assumed dust temperatures, inclination, and the dust opacity law.  
We show that the combination of these effects can lead to disc mass 
underestimates by up to factors of 2--3 at millimeter wavelengths and up to an 
order of magnitude or larger at submillimeter wavelengths.
The optically thin portions of protostellar discs are generally cooler in the 
Class I stage than the Class 0 stage since Class I discs are typically larger 
and more optically thick, and thus more shielded.  The observed disc mass 
distribution closely resembles the intrinsic distribution if this effect is 
taken into account, especially at millimeter wavelengths where optical depth 
effects are minimized.
Approximately 50\%--70\% of protostellar discs observed to date with this 
method are consistent with the masses of the gravitationally unstable discs 
formed in the simulations, suggesting that at least some protostellar discs 
are likely sufficiently massive to fragment.  
We emphasize key future work needed to confirm these results, 
including assembling larger, less biased samples, and using molecular line 
observations to distinguish between rotationally supported, Keplerian discs 
and magnetically supported pseudodiscs.
\end{abstract}



\section{Introduction}\label{sec_intro}

Circumstellar discs are ubiquitous in the star formation process.  The disc 
fraction around young stars approaches 100\% in star-forming regions with 
ages less than 1 Myr, and then steadily declines to 10\% or less for regions 
10 Myr or older \citep[e.g.,][]{wyatt2008:disks}.  Disc formation is a 
natural consequence of the conservation of angular momentum in a collapsing 
system and has long been expected to form early in the protostellar stage.   
We note that, in this paper, the protostellar stage refers to the 
evolutionary stage when protostars (and their discs if they have already 
formed) are still embedded in and accreting from their parent cores 
\citep[in other words, the Class 0 and I evolutionary stages; see, e.g., ][for a recent review on evolutionary stages]{dunham2014:ppvi}.  Such 
early disc formation has been 
confirmed by \citet{terebey1984:model}, who modified the 
\citet{shu1977:sis} solution for the inside-out collapse of a singular 
isothermal sphere to include rotation.  The formation of relatively large, 
massive, Keplerian discs very early in the protostellar stage is also found by 
non-magnetic simulations of collapsing cores 
\citep[e.g.,][]{vorobyov2011:disks}.  
However, simulations of magnetized collapsing cores have found that efficient 
magnetic braking removes angular momentum from the system and may completely 
suppress disc formation in the protostellar stage 
\citep[e.g.,][]{allen2003:magneticbraking,hennebelle2008:magneticbraking,mellon2008:magneticbraking,seifried2011:magneticbraking}.  Many potential solutions 
have been proposed to reduce the effects of magnetic braking and allow discs 
to form, including magnetic flux loss through various mechanisms, 
non-ideal MHD effects, outflow-induced envelope clearing, turbulence, and 
misalignment between the magnetic field and rotation axis 
\citep[e.g.,][]{mellon2008:magneticbraking,machida2011:magneticbraking,seifried2012:magneticbraking,li2013:misalignment,joos2012:misalignment,krumholz2013:misalignment,machida2014:disks}.  
The latter has recently been tested by two observational studies which reached 
conflicting results on the alignment (or lack thereof) between magnetic 
fields and rotation axes \citep{hull2013:misalignment,chapman2013:misalignment}.

While discs must form at some point, how early they form and how quickly they 
grow in mass and size remain key open questions.  Observations of discs 
surrounding embedded protostars are difficult because the discs are deeply 
embedded in their parent cores, and are especially challenging in the Class 
0 stage when most of the system mass is still in the core.  One of the earliest 
studies of a Class 0 protostellar disc was presented by 
\citet{harvey2003:b335}, who modeled interferometer millimeter continuum 
observations of B335 and identified the presence of a compact component 
with a mass of 0.004 \msun, which they inferred to be a disc.  
Similar inferences of Class 0 discs through analysis and modeling of compact 
components in interferometer (sub)millimeter continuum observations have been 
presented by other authors 
\citep[e.g.,][]{looney2003:disks,chiang2012:l1157,zapata2013:iras16293}, 
although the results are model-dependent and the samples are too small to 
draw any statistical conclusions.  
Two recent surveys have used larger samples (10 -- 20 protostars) to argue 
that discs form very early in the Class 0 stage 
\citep{jorgensen2009:prosac,enoch2011:disks}, although both studies reach such 
sample sizes by using very 
simple methods for separating disc and core emission.  On the other hand, 
\citet{maury2010:pdbi} found no evidence for large discs with sizes greater 
than $\sim$100 AU in an interferometer millimeter continuum survey of five 
Class 0 protostars and argued that their data were inconsistent with 
simulations that neglected magnetic braking.  However, they did not directly 
address the nature of their continuum detections, in particular whether they 
arise from discs or inner, dense cores.  Only three Class 0 
systems have confirmed detections of discs through direct detection of 
Keplerian velocity profiles in millimeter spectral line observations: L1527 
\citep{tobin2012:l1527,tobin2013:l1527}, VLA 1623A 
\citep{murillo2013:vla1623,murillo2013:vla1623a}, 
and R CrA IRS7B \citep{lindberg2014:rcra}.  
Taken together, these studies emphasize that the properties and even the 
existence of embedded protostellar discs remains quite uncertain, especially 
at the youngest (Class 0) stages of protostellar evolution.

The formation time and early evolution of circumstellar discs has several 
consequences for both star and planet formation.  If discs do indeed form early 
in the embedded stage, mass from the infalling core 
can pile up in them until they become 
gravitationally unstable and fragment.  These fragments can be driven onto 
the star through torques associated with spiral arms in the discs, causing 
short accretion bursts and a general cycle of episodic protostellar mass 
accretion 
\citep[e.g.,][]{vorobyov2005:bursts,vorobyov2006:bursts,vorobyov2010:bursts}.  
Such bursts can have significant effects on the chemistry of the surrounding 
core and planet-forming disc 
\citep[][]{lee2007:chemistry,kim2011:cb130,kim2012:co2,visser2012:chemistry,vorobyov2013:chemistry}.  
Additionally, the large changes in accretion rates lead to large changes in 
accretion luminosity, and the episodic mass accretion process driven by disc 
fragmentation predicted by \citet{vorobyov2010:bursts} provides a viable 
solution to the protostellar luminosity problem 
\citep{dunham2010:evolmodels,dunham2012:evolmodels}, whereby 
protostars are underluminous compared to simple theoretical expectations 
\citep{kenyon1990:luminosities,kenyon1994:luminosities,kenyon1995:luminosities}.  
Moreover, fragments can be ejected from discs into the intracluster
medium, producing a population of proto-brown dwarfs and very-low-mass protostars \citep{BV2012},
or settle onto stable orbits, giving rise to wide-separation giant planet and brown dwarf
companions \citep{VB2010gp,Vorob2013}.
Finally, solid cores forming in the fragment interiors may provide an alternative scenario
for the icy and terrestrial planet formation \citep{Nayakshin2010}.

Clearly, the significance of all of these effects due to disc fragmentation 
will depend on when exactly, if at all, 
discs with sufficient mass to fragment actually 
form.  While observational constraints are currently quite limited, as 
discussed above, the largest surveys to date of protostellar discs 
\citep{jorgensen2009:prosac,enoch2011:disks} depend 
on very simple assumptions to separate disc and core emission and derive 
disc masses.  If these methods successfully recover disc masses they offer a 
promising method of assessing the formation and early evolution of protostellar 
discs for statistically significant samples.  
In this paper we use synthetic observations of simulated 
protostellar core+disc systems to evaluate whether or not these assumptions 
successfully recover disc masses, and, based on those results, assess 
whether protostellar discs are sufficiently massive to fragment given the 
current observational constraints.  The organization of this paper is as 
follows:  We describe the coupling between the hydrodynamic simulations and 
radiative transfer models used to generate the synthetic observations in 
\S \ref{sec_model}, we evaluate the extent to which simple assumptions can 
be adopted to recover intrinsic disc masses from our synthetic 
interferometric observations in \S \ref{sec_results}, we discuss the 
implications and limitations of our results in \S \ref{sec_discussion} and 
\S \ref{sec_limit}, respectively, and we summarize our results in 
\S \ref{sec_summary}.

\section{Description of the Model}\label{sec_model}

To assess whether or not recent interferometer surveys of protostars are 
capable of successfully recovering the intrinsic masses of protostellar discs, 
we use models that are based on a coupling of the two-dimensional, 
numerical hydrodynamical simulations of collapsing cores presented by 
\citet{vorobyov2010:bursts} with the two-dimensional, evolutionary radiative 
transfer models of collapsing cores presented by 
\citet{dunham2010:evolmodels}.  To briefly summarize, the 
\citet{vorobyov2010:bursts} hydrodynamical simulations follow for several Myr 
the collapse of a non-magnetized 
cloud core with a fixed initial mass and initial angular 
momentum from the starless phase, through the embedded phase, and into the T 
Tauri phase, with a self-consistent calculation of the instantaneous accretion 
rates onto the disc and protostar and thus the masses of the core, disc, and 
protostar as a function of time.  The \citet{dunham2010:evolmodels} 
evolutionary models use the two-dimensional, axisymmetric, Monte Carlo 
dust radiative transfer code RADMC
\citep{dullemond2000:radmc,dullemond2004:radmc} 
to calculate the dust temperature profiles, including both internal heating 
from the protostar and external heating from the interstellar radiation field.  
Accompanying routines are then used to generate spectral energy distributions 
(SEDs) and images through ray-tracing and to calculate the complex 
interferometer visibilities.  We adopt the dust opacities of 
\citet{ossenkopf1994:oh5} 
appropriate for thin ice mantles after $10^5$ yr of coagulation at a gas 
density of $10^6$ cm$^{-3}$ 
(OH5 dust; see \S \ref{sec_results_opacity} for a discussion 
of the effects of grain growth in discs on the opacity law and resulting 
disc mass calculations), and include 
isotropic scattering off dust grains.  SEDs and complex interferometer 
visibilities at each timestep are calculated for nine different inclinations 
ranging from $i=5-85$\degree\ in steps of 10\degree, 
where $i=0$\degree\ corresponds to a pole-on (face-on) system and 
$i=90$\degree\ corresponds to an edge-on system.

\begin{table*}
\begin{center}
\caption{Model Parameters}
\label{tab_models}
\begin{tabular}{lcccccc}
\hline \hline
 & $\Omega_0$ & $r_0$ & $\Sigma_0$ & Core Outer Radius & Initial Core Mass & $\beta$\\
Model & (rad s$^{-1}$) & (AU) & (g~cm$^{-2}$) & (pc) & (\msun) &  \\
\hline
1 & 4.4\ee{-14} & 2057 & 0.06 & 0.06 & 0.922 & $5.75\times10^{-3}$ \\
2 & 4.6\ee{-13} & 343 & 0.36 & 0.01 & 0.153 & $1.74\times10^{-2}$ \\
\hline
\end{tabular}
\end{center}
\end{table*}

\subsection{Hydrodynamical Simulations}\label{sec_simulations}

Our numerical hydrodynamics model for the formation and evolution
of a young stellar object is described in detail 
in \citet{vorobyov2010:bursts} and \citet{vorobyov2013:chemistry}. 
Here, we briefly review the main concepts. The model 
includes a protostar, described by a stellar evolution
code, and a protostellar disc plus infalling envelope, both
described by a numerical hydrodynamics code. Both codes are coupled in real time,
but due to computational constraints the stellar evolution code is invoked only
every 25~yr to update the properties of the forming star.
We use the thin-disc approximation complemented by a calculation
of the vertical scale height in both the disc and envelope
determined in each computational cell using an assumption
of local hydrostatic equilibrium. The resulting model has a flared
structure with the vertical scale height increasing with radial distance.  
Both the disc and envelope receive a fraction of the irradiation energy from 
the central protostar. 

The main physical processes
taken into account when computing the evolution of the
disc and envelope include viscous and shock heating, irradiation
by the forming star, background irradiation, radiative cooling
from the disc surface, and self-gravity. The corresponding
equations of mass, momentum, and energy transport are
\begin{equation}
\label{cont}
\frac{{\partial \Sigma }}{{\partial t}} =  - \nabla_p  \cdot 
\left( \Sigma \bl{v}_p \right),  
\end{equation}
\begin{eqnarray}
\label{mom}
\frac{\partial}{\partial t} \left( \Sigma \bl{v}_p \right) &+& \left[ \nabla \cdot \left( \Sigma \bl{v_p}
\otimes \bl{v}_p \right) \right]_p =   - \nabla_p {\cal P}  + \Sigma \, \bl{g}_p + \\ \nonumber
& + & (\nabla \cdot \mathbf{\Pi})_p, 
\label{energ}
\end{eqnarray}
\begin{equation}
\frac{\partial e}{\partial t} +\nabla_p \cdot \left( e \bl{v}_p \right) = -{\cal P} 
(\nabla_p \cdot \bl{v}_{p}) -\Lambda +\Gamma + 
\left(\nabla \bl{v}\right)_{pp^\prime}:\Pi_{pp^\prime}, 
\end{equation}
where subscripts $p$ and $p^\prime$ refers to the planar components $(r,\phi)$ 
in polar coordinates, $\Sigma$ is the mass surface density, $e$ is the internal energy per 
surface area, 
${\cal P}$ is the vertically integrated gas pressure calculated via the ideal equation of state 
as ${\cal P}=(\gamma-1) e$ with $\gamma=7/5$,
$\bl{v}_{p}=v_r \hat{\bl r}+ v_\phi \hat{\bl \phi}$ is the velocity in the
disc plane, and $\nabla_p=\hat{\bl r} \partial / \partial r + \hat{\bl \phi} r^{-1} 
\partial / \partial \phi $ is the gradient along the planar coordinates of the disc. 
The gravitational acceleration in the disc plane, $\bl{g}_{p}=g_r \hat{\bl r} +g_\phi \hat{\bl \phi}$, takes into account self-gravity of the disc, found by solving for the Poisson integral, 
and the gravity of the central protostar. 
Turbulent viscosity is taken into account via the viscous stress tensor 
$\mathbf{\Pi}$. We parameterize the magnitude of kinematic viscosity $\nu$ using 
the $\alpha$-prescription with a spatially and temporally uniform $\alpha=5\times 10^{-3}$.
The radiative cooling $\Lambda$ in equation~(\ref{energ}) is determined using the diffusion
approximation of the vertical radiation transport in a one-zone model of the vertical disc 
structure \citep{Johnson2003}, while the radiative heating $\Gamma$
is calculated using the irradiation temperature at the disc surface determined
by the stellar and background black-body irradiation. For more details the reader
is referred to \citet{vorobyov2010:bursts} and \citet{vorobyov2013:chemistry}.

For the initial gas surface density $\Sigma$  and angular velocity $\Omega$ distributions
we take those typical of pre-stellar cores formed as a result of the slow expulsion 
of magnetic field due to ambipolar diffusion \citep{Basu1997}
\begin{equation}
\Sigma={r_0 \Sigma_0 \over \sqrt{r^2+r_0^2}}\:,
\label{dens}
\end{equation}
\begin{equation}
\Omega=2\Omega_0 \left( {r_0\over r}\right)^2 \left[\sqrt{1+\left({r\over r_0}\right)^2
} -1\right].
\label{omega}
\end{equation}
Here, $\Omega_0$ and $\Sigma_0$ are the angular velocity and gas surface
density at the center of the core and $r_0 =\sqrt{A} c_{\rm s}^2/\pi G \Sigma_0 $
is the radius of the central plateau, where $c_{\rm s}$ is the initial sound speed in the core. 
The gas surface density distribution described by equation~(\ref{dens}) can
be obtained (to within a factor of unity) by integrating the 
three-dimensional gas density distribution characteristic of 
Bonnor-Ebert spheres with a positive density-perturbation amplitude A \citep{Dapp2009}.
The value of $A$ is set to 1.2 and the initial gas temperature is set to 10~K.

\begin{figure*}
\centering
\includegraphics[width=16cm]{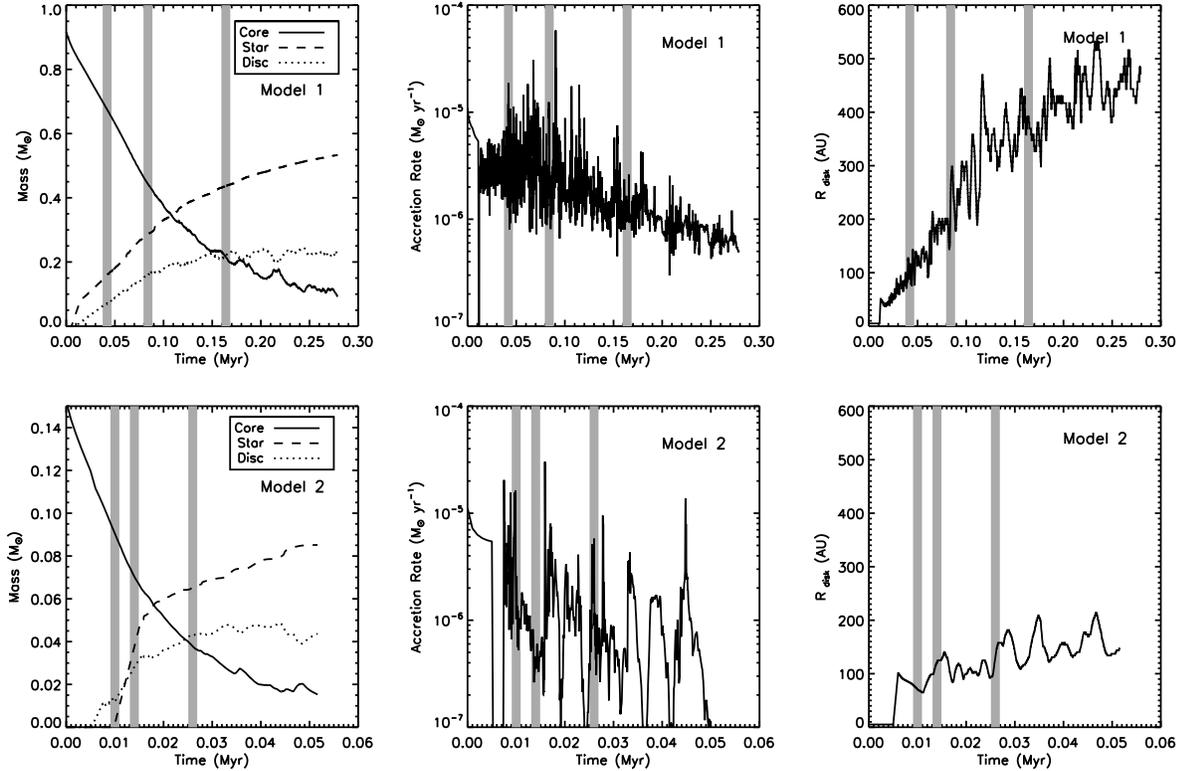}
\caption{{\it Left Panels:} Time evolution of the core, star, and disc masses 
for models 1 (top) and 2 (bottom).  {\it Middle Panels:} Time evolution of the 
protostellar mass accretion rates for models 1 (top) and 2 (bottom).  
{\it Right Panels:} Time evolution of the disc radius for models 1 (top) and 
2 (bottom).  In each 
panel the thick gray lines mark, from left to right, the Class 0, Class 0/I, 
and Class I timesteps considered in this paper (see \S \ref{sec_results} for 
details).}
\label{fig_models}
\end{figure*}

In this paper two different models with different initial conditions 
are coupled with the radiative transfer models, as described below in 
\S \ref{sec_radtrans}.  Table \ref{tab_models} lists the initial conditions 
for each model, including the initial central angular velocity ($\Omega_0$), 
the initial outer radius of the core ($r_0$), the initial central surface 
density ($\Sigma_0$), the radius of the central region with an initially flat 
surface density profile ($r_0$), the initial core 
mass, and the initial ratio of rotational to gravitational energy ($\beta$).  
Figure \ref{fig_models} plots the time evolution of the core, star, and 
disc masses, protostellar mass accretion rates, and disc radii for each model.  
These two models are chosen to represent opposite extremes: model 1 has a 
near-solar mass core that forms a sub-solar mass star, whereas model 2 
initially has a much lower core mass, giving birth to a very-low mass star.   
As a consequence, model 1 forms a much larger, more massive disc than model 2 
\citep[see Figure \ref{fig_models} and][for details]{vorobyov2010:bursts}.
Nevertheless, both discs are gravitationally unstable and prone to 
fragmentation, owing to rather high initial values of $\beta$, and also reveal 
highly variable accretion rates caused by disc gravitational instability. 

\begin{figure}
\resizebox{\hsize}{!}{\includegraphics{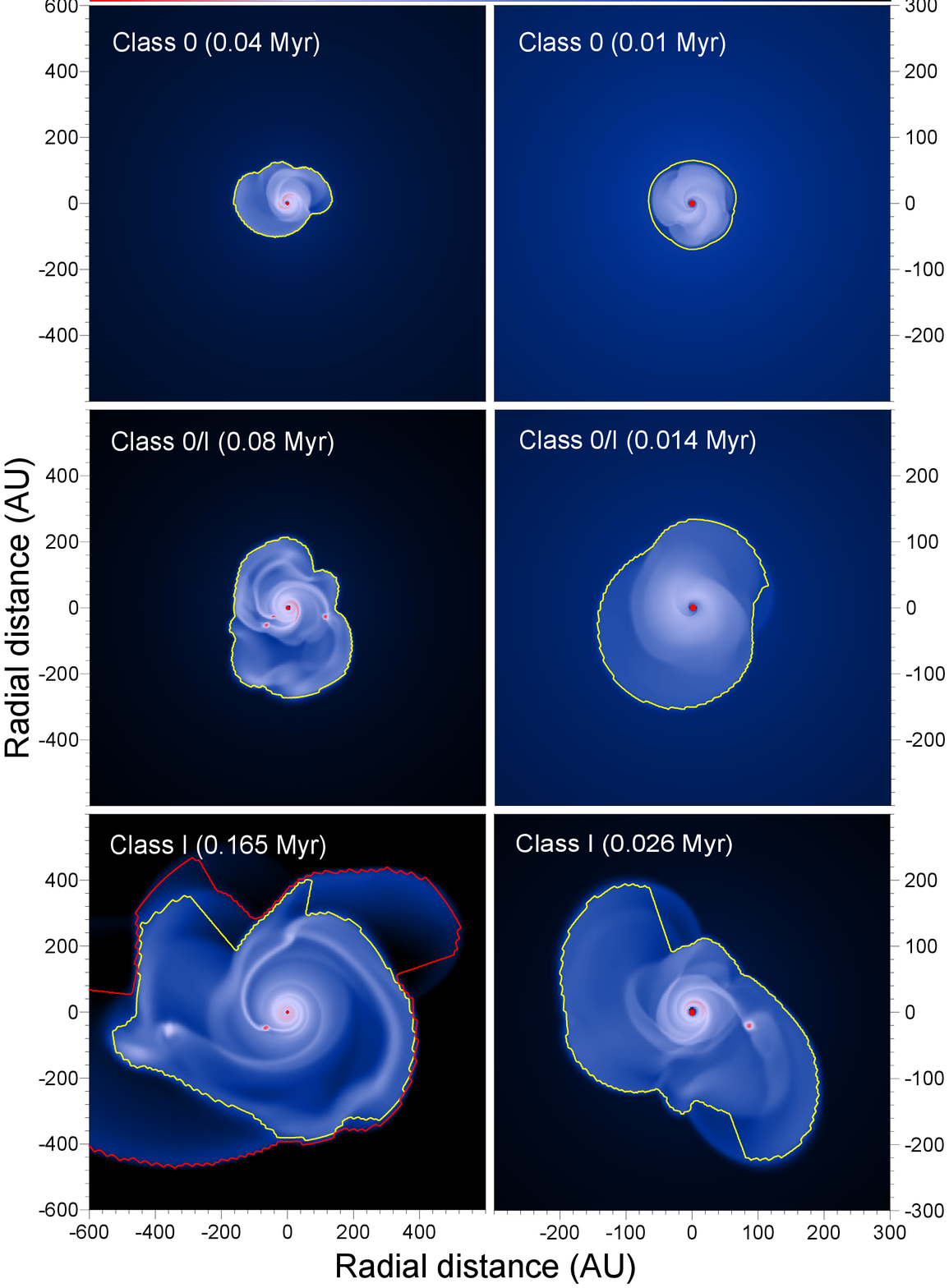}}
\caption{Gas surface density images (in~log~g~cm$^{-2}$)
in model~1 (left column) and model~2 (right column) in the 
Class 0, Class 0/I, and Class I timesteps considered in this paper (see 
\S~\ref{sec_results} for details). The time in parentheses is counted from the 
formation of the first hydrostatic core.  The yellow lines outline the discs 
according to the adopted critical surface density of 
$\Sigma_{\rm cr}=0.5$~g~cm$^{-2}$.  The red line in the lower left panel 
outlines the disc assuming a lower $\Sigma_{\rm cr}$ of 0.1~g~cm$^{-2}$ (see 
text for details).  Fragments forming in the disc in the Class 0/I and I 
stages are clearly visible.}
\label{fig_image}
\end{figure}

Figure~\ref{fig_image} shows the gas surface density distribution 
in model~1 (left column) and model~2 (right column) in the 
Class 0, Class 0/I, and Class I timesteps considered in this paper (see 
\S \ref{sec_results} for details). For both models, only the inner several 
hundred AU are shown; the full computational area is approximately 10 times 
larger.  The yellow lines outline the disc in each stage as identified using 
the algorithm described in \citet{dunham2012:evolmodels}, which is based on a 
critical surface density for the disc-to-envelope transition of 
$\Sigma_{\rm cr}=0.5$~g~cm$^{-2}$.  We found that this algorithm works well in 
the Class 0/I stage, but may somewhat overestimate the disc mass in the Class 
0 stage (when the inner part of the infalling envelope may have densities 
exceeding $\Sigma_{\rm cr}$) and somewhat underestimate the disc mass in the 
Class I stage (when the disc spreads out and its density drops below 
$\Sigma_{\rm cr}$ in the outer parts). For instance, a lower value of 
$\Sigma_{\rm cr}=0.1$~g~cm$^{-2}$ would result in larger disk sizes, as 
illustrated by the red curve in the bottom-left panel of 
Figure~\ref{fig_image}. However, the net increase in the disc mass is only 
$4\%$, owing to a rather low density at the disc periphery.

\subsection{Radiative Transfer Calculations}\label{sec_radtrans}

We couple the simulations described above with radiative transfer models 
to calculate the time evolution of the SEDs of cores collapsing according 
to the predictions of the simulations, as well as images and interferometer 
visibilities at all wavelengths.  
The coupling between the simulations and radiative transfer models was first 
described by \citet{dunham2012:evolmodels}.  We give a brief overview here but 
refer to \citet{dunham2012:evolmodels} for more 
details.  Since the simulations do not provide the full volume density 
structure, in order to set up the physical structure 
of the radiative transfer models we take analytic profiles for the core and 
disc density structure and re-scale them at each timestep according to known 
parameters from the simulations (see \S \ref{sec_limit_structure} for a 
discussion of the limitations of this approach and prospects for future 
improvements).  These parameters include the initial core 
mass (\mcore) and radius (\renv),
along with the time evolution of the core mass, disc mass (\mdisc),
protostellar mass (\mstar), disc outer radius (\rdisc), accretion rate onto
the protostar (\mdotstar), and accretion rate onto the disc (\mdotdisc).  

For the core, we adopt the density profile given by the 
\citet{terebey1984:model} solution for the collapse of a slowly rotating core.  
This solution gives a core that is initially a spherically symmetric, singular 
isothermal sphere with a density distribution $n \propto r_{core}^{-2}$, 
identical to the classic \citet{shu1977:sis} solution.  As collapse proceeds, 
the solution takes on two forms: an outer solution that is similar to the 
non-rotating, spherically symmetric solution and an inner solution that 
exhibits flattening of the density profile.  This model is parameterized by the 
initial angular velocity of the core and the time since the formation of the 
protostar.  At each timestep we truncate the solution at the given initial 
\renv\ for each model and 
then renormalize the density profile so that the core mass matches that given 
by the simulations.  Once the material at the outer core boundary beings 
collapsing, which occurs once the infall 
radius\footnote{The infall radius is the radius within which the core is 
collapsing.  It starts at the center and moves outward at the sound speed.} 
exceeds the initial outer radius, we use the velocity profiles given by the 
\citet{terebey1984:model} solution to allow the core radius to decrease.

The disc structure follows a power law in the 
radial coordinate and a Gaussian in the vertical coordinate, with the density 
profile given by:
\begin{equation}\label{eq_disc_density_profile}
\rho_{disc}(s,z) = \rho_0 \left(\frac{s}{s_o} \right)^{-n} 
exp\left[ -\frac{1}{2} \left(\frac{z}{H_s}\right)^2 \right] \qquad ,
\end{equation}
where $z$ is the distance above the midplane ($z=rcos\theta$, with $r$ and 
$\theta$ the usual radial and zenith angle spherical coordinates), $s$ is the 
distance in the midplane from the origin ($s= \sqrt{r^2 - z^2}$), and 
$\rho_0$ is the density at the reference midplane distance $s_0$.  
The quantity $H_s$ is the disc scale height and is given by 
$H_s = H_o \left(\frac{s}{s_o}\right)^\gamma$, where $H_o$ is the scale height 
at $s_o$.  The parameter $\gamma$ describes how the scale height changes with 
$s$ and sets the flaring of the disc.  We set $H_o = 10$ AU at $s_o = 100$ AU 
and $\gamma = 1.25$ (see \citet{dunham2010:evolmodels} and references therein 
for a more detailed discussion of these quantities), giving flared discs very 
similar to those found in the hydrodynamical simulations (see Figure 11 of 
\citet{vorobyov2010:bursts}).  The disc surface density profile, calculated by 
integrating Equation \ref{eq_disc_density_profile} over the vertical 
coordinate $z$, has a radial power-law index of $\Sigma(s) \propto s^{-p}$, 
where $p = n - \gamma$. Unlike our previous work 
\citep{dunham2010:evolmodels} where we set $p$ to a fixed value, 
here we determine $p$ by fitting a power-law to 
the azimuthally-averaged surface density profile given by the 
simulations, and use this to set $n$ 
($p=-1.68, -1.44$ for the Class 0 and I evolutionary stages of model 1, 
respectively, and $p=-1.77, -1.36$ for the Class 0 and I evolutionary stages 
of model 2, respectively).  Finally, the disc inner 
radius is set equal to the dust destruction radius, calculated 
(assuming spherical, blackbody dust grains) as
\begin{equation}
R_{disc}^{in} = \sqrt{\frac{L_*}{16 \pi \sigma T_{dust}^4}} \qquad ,
\end{equation}
where $L_*$ is the protostellar luminosity (see below) and $T_{dust}$ is the 
dust destruction temperature 
\citep[assumed to be 1500 K; e.g.,][]{cieza2005:tdust}.

The total internal luminosity of the protostar and 
disc at each point in the collapse from core to star contains six components:  
(1) luminosity arising from accretion from the core directly onto the 
protostar, (2) luminosity arising from accretion from the core onto the disc, 
(3) luminosity arising from accretion from the disc onto the protostar, (4) 
disc ``mixing luminosity'' arising from luminosity released when newly 
accreted material mixes with existing disc material, (5) luminosity arising 
from the release of energy stored in differential rotation of the protostar, 
and (6) photosphere luminosity arising from gravitational contraction and 
deuterium burning.  The first five components are calculated following 
\citet{adams1986:protostars}, using direct inputs from the simulations for 
the accretion rates, masses, and sizes of the protostar, disc, and core 
(further details can be found in \citet{young05:evolmodels}, 
\citet{dunham2010:evolmodels}, and \citet{dunham2012:evolmodels}).  
The photospheric luminosity and protostellar radius are 
calculated from the stellar evolution code
of \citet{BC2010} coupled to the main numerical hydrodynamic code 
(see \citet{vorobyov2013:chemistry} 
for details).  
While the total amount of accretion luminosity is correct, 
we do not include viscous disc heating since the radiative transfer code can 
only incorporate heating due to one central, internal source of photons and 
an external radiation field.  Viscous heating is only significant within the 
inner $20-30$ AU of protostellar discs, and even in those locations it only 
increases the disc temperatures by much less than a factor of two (see, e.g., 
Figure 4 of \citet{vorobyov2010:disks}).  Since the discs we consider are 100 
AU or larger, even at the earliest times (see right panels of Figure 
\ref{fig_models}), small deviations to warmer temperatures in the inner discs 
than calculated from our radiative transfer models will have a negligible 
impact on our results.

Finally, we also include external luminosity arising from heating of the core 
by the Interstellar Radiation Field (ISRF).  As in our previous papers, we 
adopt the \citet{black1994:isrf} ISRF, modified in the ultraviolet to 
reproduce the \citet{draine1978:isrf} ISRF, and then extincted by $A_V = 0.5$ 
of dust with properties given by \citet{draine1984:dust} to simulate extinction 
by the surrounding lower density environment.

\section{Results}\label{sec_results}

The two largest surveys of protostellar discs to date, and the only two surveys 
to include Class 0 protostars, are those presented by 
\citet{jorgensen2009:prosac} and \citet{enoch2011:disks}.  
Both studies were based on 
measuring the total 1.1 or 1.3 mm interferometric amplitude at baselines of 
50 k$\lambda$ for protostellar sources at distances ranging from 125 to 
415 pc.  At these distances, emission on baselines of 50 \kl\ 
corresonds to spatial scales of approximately $600-2000$ AU, and the basic 
assumption made is that this emission has completely resolved out 
the surrounding core but not yet begun to resolve out the disc.  Under this 
assumption, and further assuming that the observed emission is from 
optically thin, isothermal dust at a temperature of 30 K, disc masses can then 
be calculated assuming a standard gas-to-dust ratio of 100.  A number of open 
questions surrounding these assumptions must be answered before any general 
conclusions can be drawn about the formation and early evolution of 
protostellar discs, including:

\begin{enumerate}
\item Is it true that the total amplitude measured at baselines of 50 
k$\lambda$ is free of contamination from the surrounding core and recovers 
the total emission from the disc, and how do these answers depend on the 
distance to the source?
\item Is it valid to assume that the (sub)millimeter emission is optically 
thin, or is mass missed due to optically thick emission?
\item Is 30 K the best temperature to assume?
\item How do the answers to all of these questions depend on the source 
inclination and the wavelength of the observations?
\end{enumerate}

While \citet{jorgensen2009:prosac} partially addressed the 
first question by attempting to estimate and remove contaminating emission from 
the surrounding core, it is unclear how accurate their corrections are since 
they used simple, one-dimensional models to derive them.  
To answer these questions, we take each of the models listed in Table 
\ref{tab_models} and, for source distances ranging from 100 pc to 500 pc in 
steps of 50 pc, calculate the complex interferometer visibilities at four 
wavelengths: 3.4 mm, 1.3 mm, 0.7 mm (700 \um), and 0.35 mm (350 \um) 
(approximately 90, 230, 430, and 850 GHz, respectively).  The visibilities 
are calculated using an accompanying routine in the RADMC package and are 
generated assuming the Atacama Large Millimeter Array (ALMA) 
primary beam at each of the wavelengths listed 
above, for nine inclinations ranging from 5\degree\ to 85\degree\ in steps of 
5\degree.  The contributions to the visibility amplitudes from the 
core and disc are calculated separately by coupling the hydrodynamic 
simulations with the radiative transfer models a second time with the disc 
mass set to zero, 
generating a set of core-only visibilities, and then subtracting their 
amplitudes from the combined core+disc amplitudes to generate a set of 
disc-only amplitudes.  We 
consider three timesteps for each model: one early in the Class 0 stage when 
only 25\% of the initial core mass has accreted onto 
the protostar and/or disc, one at the Class 0/I boundary when 50\% of the 
initial core mass has accreted onto the protostar and/or disc, and one late 
in the Class I stage when 75\% of the initial core mass has accreted onto 
the protostar and/or disc.  While we caution that there is not always a 
one-to-one correspondence between observational Class and physical Stage 
due to the effects of geometry and episodic accretion 
\citep[e.g.,][]{dunham2014:ppvi}, for simplicity here we use the term Class 
to refer to both.  In the following subsections we use these 
synthetic observations to answer the questions listed above.

\subsection{Contamination From the Surrounding Core}\label{sec_results_core}

\begin{figure}
\resizebox{\hsize}{!}{\includegraphics{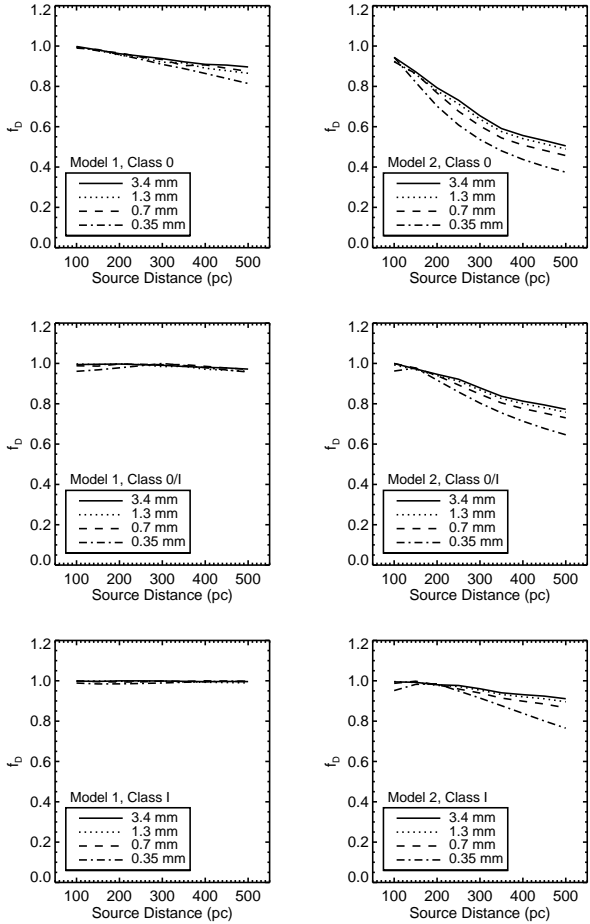}}
\caption{$f_D$, the ratio of the amplitude at 50 \kl\ in the disc-only model 
to the amplitude at 50 \kl\ in the disc+core model, plotted as a function 
of distance to the source for distances ranging from 100 to 500 pc.  The 
amplitudes are calculated for a source inclination angle of 45\degree.}
\label{fig_fd}
\end{figure}

To evaluate whether or not the total amplitude measured at a baseline of 50 
\kl\ is free of contamination from the surrounding core and truly 
represents emission from only the disc, we define the quantity $f_D$ as 
the ratio of the amplitude at 50 \kl\ in the disc-only visibilities to the 
amplitude at 50 \kl\ in the combined (core+disc) visibilities.  Figure 
\ref{fig_fd} plots $f_D$ versus distance to the source for each timestep of 
each model for each of the four wavelengths considered in this paper, 
at an inclination angle of 45\degree.  For 
model 1, which has a relatively massive disc, there is almost no contamination 
from the core.  Even during the Class 0 stage when the core still contains 
the majority of the system mass, at the largest distances where the core is 
the least well-resolved, at most only 20\% of the amplitude measured at 
50 \kl\ arises from the core.

As seen in Figure \ref{fig_fd}, 
the contamination level is somewhat higher for model 2, which is not 
surprising given that this model forms a less massive disc.  During the Class 0 
stage, between 10\% -- 60\% of the observed amplitude at 50 \kl\ arises 
from the surrounding core, with the highest contamination levels occuring 
for submillimeter observations of sources at the largest distances.  At 
the Class 0/I boundary the contamination levels are lower, in the range of 
0\% -- 35\%, and they are even lower in the Class I stage, ranging between 
0\% and 20\%.  These results for both models do not show any significant 
dependence on inclination angle.

Overall, we find that the surrounding core does not significantly contaminate 
the visibility amplitudes at 50 \kl.  For cores that form relatively massive 
discs the contamination is almost completely negligible for all but the 
earliest times and largest source distances, where it can reach up to 20\%.  
For cores that form less massive discs, up to $\sim$60\% of the amplitude 
may actually arise from the core rather than the disc for Class 0 sources 
at distances of 400 -- 500 pc observed at submillimeter wavelengths.  For 
any combination of closer distances, more evolved sources, and longer 
wavelength observations the contamination is less.  
These results are in general agreement 
with those of \citet{jorgensen2009:prosac}, who used simple, one-dimensional 
models to estimate that cores contribute about 10\% -- 30\% of the observed 
interferometer amplitudes at 50 \kl.

\subsection{Resolving Out the Disc}\label{sec_results_resolve}

Depending on both the size of the disc and the distance to the source, the 
interferometer amplitudes at 50 \kl\ may not trace the full mass of the disc 
if the disc is partially resolved.  Indeed, \citet{enoch2011:disks} noted 
that, due to the disc being partially resolved out at 50 \kl, the inferred 
disc mass for one source in their sample 
was 30\% lower than the mass obtained from detailed radiative 
transfer modeling performed in an earlier study \citep{enoch2009:disks}.  
Since the \citet{vorobyov2010:bursts} 
simulations form relatively large, massive discs, we evaluate here whether 
disc masses calculated from the amplitues at 50 \kl\ are underestimated due 
to the discs being partially resolved out.  For each timestep of each 
model we define the quantity $f_{TD}$ as the ratio of amplitudes in the 
disc-only model at 50 \kl\ to that at 0 \kl.  In this sense $f_{TD}$ measures 
the fraction of total disc flux recovered at 50 \kl.

\begin{figure}
\resizebox{\hsize}{!}{\includegraphics{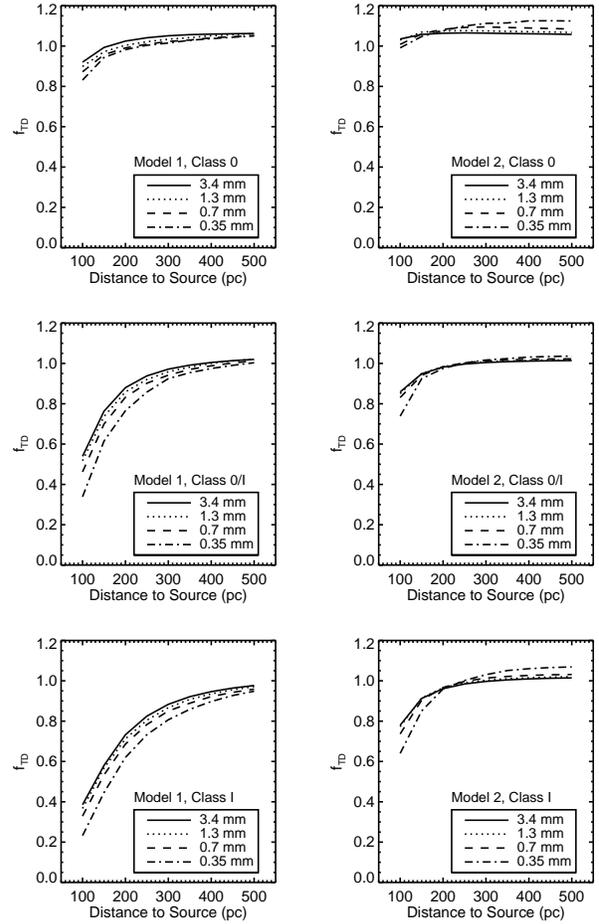}}
\caption{$f_{TD}$, the ratio of the amplitude at 50 \kl\ in the disc-only 
model to the total amplitude of the disc-only model (the amplitude at 0 \kl), 
plotted as a function of distance to the source for distances ranging from 100 
to 500 pc.  The amplitudes are calculated for a source inclination angle of 
45\degree.}
\label{fig_ftd}
\end{figure}

Figure \ref{fig_ftd} plots $f_{TD}$ versus distance to the source for each 
timestep of each model for each of the four wavelengths considered in this 
paper, at an inclination angle of 45\degree\ (the choice of inclination 
angle is not found to have any significant effect on the results).  In the 
Class 0 stage the amplitudes at 50 \kl\ do recover essentially all of the 
disc emission except for sources closer than 200 pc, where up to 20\% of the 
emission is resolved out in models with larger and more massive discs.  
However, as the disc grows larger at later evolutionary stages, more of 
the disc emission is resolved out.  For model 2, the disc remains small 
enough such that essentially no emission is resolved out for source distances 
greater than $\sim$ 200 pc; for closer sources, up to 20\% is resolved out 
at the Class 0/I boundary, and up to 30\% late in the Class I stage.  
For model 1, on the other hand, the disc becomes partially resolved out 
for all but the largest source distances.  Up to 50\% of the total disc 
emission is resolved out at the Class 0/I boundary for observations at 
3.4 and 1.3 mm (and even more at submillimeter wavelengths), and up to 60\% 
late in the Class I stage (again, even more at submillimeter wavelengths).

Combining the results of the last two sections, interferometer amplitudes at 
50 \kl\ will overestimate the emission from a disc by up to about a factor of 
two in the worst cases due to contaminating emission from the surrounding 
core.  On the other hand, they will also underestimate the emission from a disc 
by a similar factor due to partially resolving out the disc, as long as the 
observations are obtained at millimeter wavelengths.  For large 
enough samples of objects over different ranges of source distances and 
intrinsic disc properties, these effects will at least partially cancel each 
other out.  For observations obtained at submillimeter wavelengths, 
however, up to 80\% of the disc emission can be resolved out, and as this 
effect is larger than the contamination from the surrounding core, 
submillimeter observations used to measure disc masses in this manner will 
systematically underestimate the true masses.

\subsection{Optical Depth, Temperature, and Inclination}\label{sec_results_mass}

\begin{figure}
\resizebox{\hsize}{!}{\includegraphics{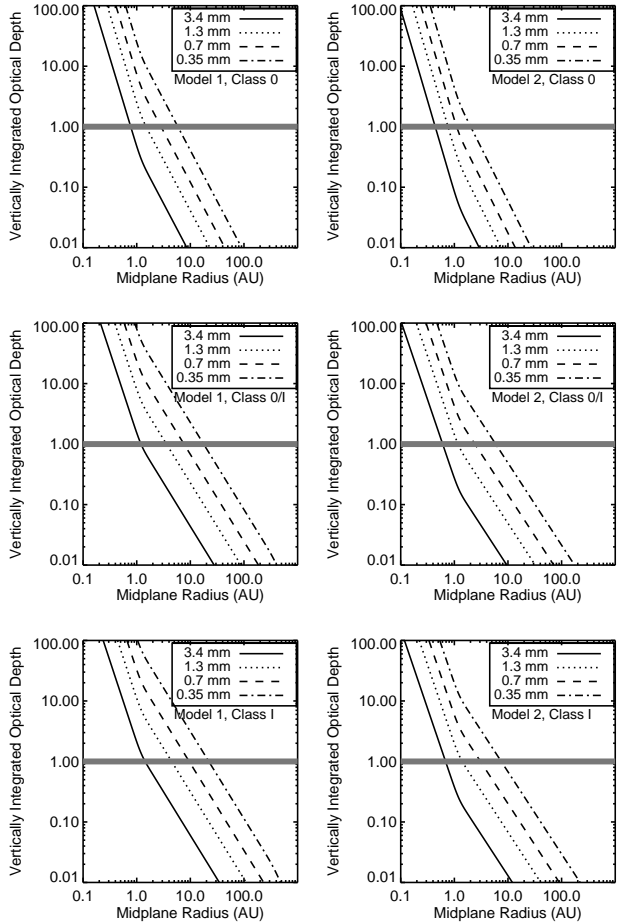}}
\caption{The vertically integrated optical depth through the disc for each 
timestep of each model, plotted as a function of the midplane radial distance.  
The optical depth is calculated and displayed for each of the four wavelengths 
considered in this paper.  The thick gray line in each panel marks the limit 
between optically thick ($\tau > 1$) and optically thin ($\tau < 1$).  The 
optical depths are calculated using OH5 opacities (see text for details).}
\label{fig_tau}
\end{figure}

In this section we evaluate the effects of optical depth, temperature, and 
inclination on our ability to recover the actual disc masses from our 
synthetic observations.  The observed disc emission will only trace the 
total mass of the disc if the emission remains optically thin throughout the 
full vertical and radial extent of the disc.  This condition is known to be 
false in older Class II discs where the inner regions are often optically 
thick \citep[e.g.,][]{andrews2005:disks}.  To evaluate whether the protostellar 
discs predicted by the \citet{vorobyov2010:bursts} simulations are also 
partially optically thick, 
Figure \ref{fig_tau} plots, for each wavelength considered in this 
paper, the vertically integrated optical depth through the disc for each 
timestep of each model as a function of the midplane radial distance.  
For self-consistency the optical depths are calculated using OH5 dust 
opacities, since these opacities were used in the radiative transfer models to 
generate the synthetic observations.

The vertically integrated optical depth at a fixed midplane radial distance 
increases with decreasing wavelength, as expected given that the opacities 
also increase with decreasing wavelength.  In all cases, the discs are not 
optically thin all the way down to the midplane for all radii.  For model 1, 
the disc emission becomes optically thin beyond a few AU for the millimeter 
wavelengths but 
only beyond ten to a few tens of AU for the submillimeter wavelengths.  
For model 2, which forms a smaller, less massive disc that is 
consequently less optically thick, the disc is generally optically thin
beyond a few AU except for the shortest wavelengths and latest evolutionary 
stages, where it is only optically thin beyond about 10 AU.  
These results demonstrate that the discs formed by the 
\citet{vorobyov2010:bursts} simulations are not fully optically thin, 
in agreement with other numerical studies (see e.g. \citet{Rice2010}).  
However, they are not optically thick enough to cause large mass 
underestimates.  For a disc with a surface density profile 
$\Sigma(s) \propto s^{-p}$, the mass contained within the midplane 
radius $s$ is $M(<s) \propto s^{2-p}$.  At each timestep the disc size (Figure 
\ref{fig_models}), the power-law index (see \S \ref{sec_radtrans}), and the 
midplane radius where the emission transitions to optically thin (Figure 
\ref{fig_tau}) can be combined to calculate the fraction of total mass 
contained within the optically thick region of the disk.  For the shortest, 
most optically thick wavelength considered here (0.35 mm), we calculate this 
fraction to be 41\%, 23\%, and 20\% for the 
Class 0, 0/I, and I timesteps of model 1, respectively, and 45\%, 14\%, and 
14\% for model 2.  Thus the total disc mass underestimates due to optically 
thick emission are less than a factor of two even at the shortest submillimeter 
wavelengths (except for edge-on lines-of-sight; see below).

With these results in mind, 
we calculate, at each wavelength, the total disc mass inferred 
from the synthetic interferometer amplitude at 50 \kl\ for each timestep of 
each model.  The mass is given by

\begin{equation}\label{eq_dustmass}
M = 100 \frac{d^2 S_{\nu}(50~\kl)}{B_{\nu}(T_D) \kappa_{\nu}} \quad ,
\end{equation}
where $S_{\nu}(50~\kl)$ is the interferometer amplitude at 50 \kl, 
$B_{\nu}(T_D)$ is the Planck 
function at the assumed isothermal dust temperature $T_D = 30$ K, 
$\kappa_{\nu}$ is the dust opacity, $d$ is the distance to the source, and 
the factor of 100 is the assumed gas-to-dust ratio.  We again adopt the OH5 
dust opacities for self-consistency with the radiative transfer models 
(see \S \ref{sec_results_opacity} for discussion of the opacities), and use 
the synthetic interferometer visibilities calculated for a source distance of 
250 pc.

\begin{figure}
\resizebox{\hsize}{!}{\includegraphics{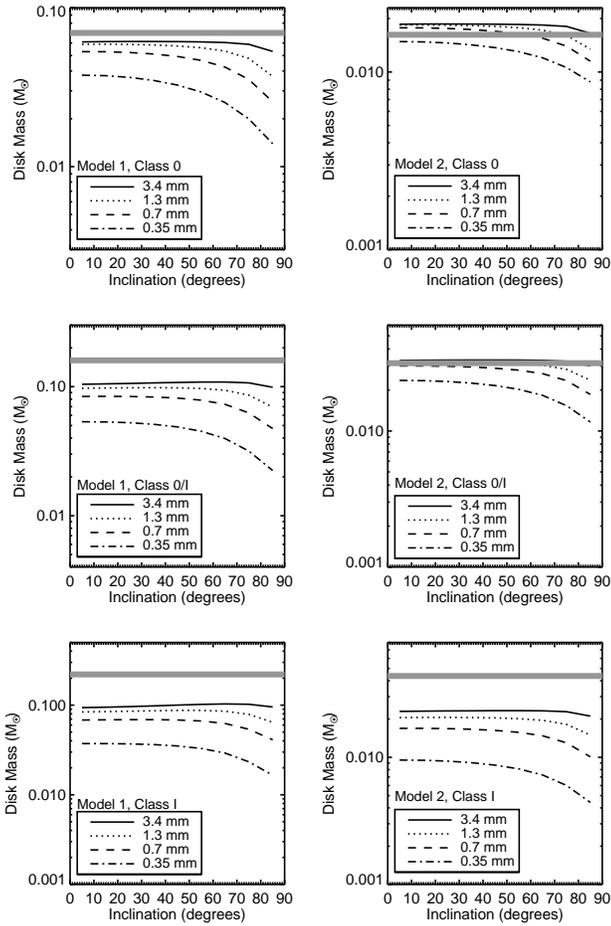}}
\caption{Disc mass calculated from the interferometer visibilities at 50 
\kl\ at each wavelength, for each timestep of each model, calculated according 
to Equation \ref{eq_dustmass} (see text for details) and plotted versus the 
inclination assumed to generate the synthetic visibilities.  In each panel the 
thick gray line shows the actual disc mass.}
\label{fig_mass}
\end{figure}

\begin{figure}
\resizebox{\hsize}{!}{\includegraphics{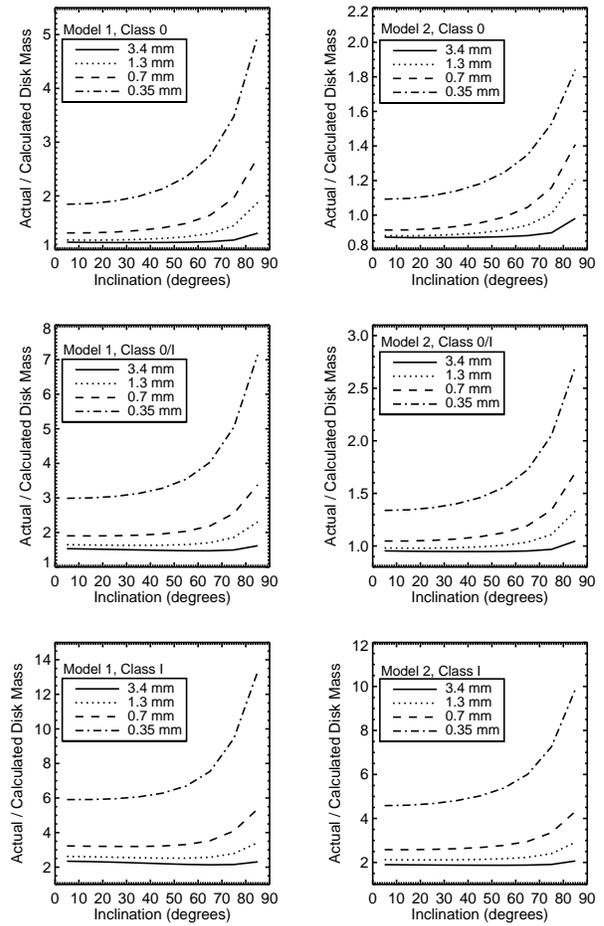}}
\caption{Ratio of actual to calculated disc mass at each wavelength, 
for each timestep of each model, plotted versus the inclination assumed to 
generate the synthetic interferometer visibilities.}
\label{fig_massratio}
\end{figure}

Figure \ref{fig_mass} plots the resulting disc masses calculated at each 
wavelength as a function of inclination, and Figure \ref{fig_massratio} 
plots the ratio of actual to calculated disc masses, again as a function of 
inclination.  
In both Figures, the calculated disc masses are seen to converge 
toward a single value for all inclinations less than $\sim$60\degree\ 
as the wavelength increases.  The increasing underestimates in disc mass at 
progressively shorter wavelengths is due to a combination of two factors: 
(1) the emission is progressively more optically thick, and (2) the 
emission is progressively farther from the Rayleigh-Jeans limit, thus 
temperature overestimates cause progressively larger disc mass underestimates.  
The importance of the latter factor is emphasized since we have already shown 
that the underestimates due to optically thick emission are less than a factor 
of two even at the shortest wavelengths.  

In four of the six panels the masses calculated from the 
longest, most optically thin wavelengths converge to lower masses than the 
actual disc masses.  These underestimates grow more significant with later 
evolutionary stages, reaching up to factors of two late in the Class I stage, 
and are due to the fact that the assumed isothermal temperature of 30 K is 
often too high.  As the discs are not truly isothermal, 
the correct temperature to assume in Equation \ref{eq_dustmass} is the 
mass-weighted average temperature of the optically thin portions of the disc.  
Table \ref{tab_temperatures} lists these temperatures for each timestep of 
each model.  The temperatures decrease as the evolutionary stages increase, 
since as the disc evolves it grows in mass and size and is optically thick 
to larger radii (see Figure \ref{fig_tau}), leading to more shielding of the 
optically thin regions.  Consequently, masses calculated 
assuming $T_D = 30$ K lead to progressively larger underestimates as 
evolutionary stage increases.  
A similar trend of decreasing disc temperatures leading 
to increasing mass underestimates with evolutionary stage was also noted by 
\citet{jorgensen2009:prosac}.  The small overestimate (less than 20\%) in the 
disc mass for the Class 0 stage of model 2 is due to the temperature being 
slightly above the assumed 30 K (35 K, see Table \ref{tab_temperatures}).

\begin{table}
\caption{Isothermal Temperatures}
\label{tab_temperatures}
\begin{tabular}{lcc}
\hline \hline
      &                    & Temperature \\
Model & Evolutionary Stage & (K)         \\
\hline
1     & 0                  & 27          \\
1     & 0/I                & 20          \\
1     & I                  & 15          \\
2     & 0                  & 35          \\
2     & 0/I                & 32          \\
2     & I                  & 17          \\
\hline
\end{tabular}
\end{table}

Assuming a random distribution of source orientations on the sky, 50\% of all 
sources will be viewed at inclinations of 60\degree\ or greater.  
For such sources, the lines-of-sight will pass partially or fully through the 
disc midplane where the densities (and thus optical depths) are the largest.  
As illustrated by Figures \ref{fig_mass} and \ref{fig_massratio}, the disc 
mass underestimates due to the combination of optically thick emission 
and incorrect isothermal dust temperatures for sources 
with inclinations greater than $\sim$60\degree\ are generally less than 50\% 
for observations at millimeter wavelengths, but can reach up to factors of 
5--10 or larger for submillimeter wavelengths.  

\subsection{Dust Opacity Law}\label{sec_results_opacity}

As noted above, the radiative transfer models used to generate the 
synthetic interferometer visibilities assumed the OH5 dust opacity law, 
so to be self-consistent we also adopted the OH5 dust opacities when 
calculating disc masses according to Equation \ref{eq_dustmass}.  
Defining $\beta$ as the power-law index of the dust opacity law 
($\kappa \propto \nu^{\beta}$, where $\kappa$ is the dust opacity at 
frequency $\nu$), optically thin emission will scale as 
$I \propto \nu^{\alpha}$, where $\alpha$ is the spectral index and is given 
by $\alpha = 2 + \beta$.  For OH5 dust, $\beta = 1.75$ over all 
submillimeter and millimeter wavelengths, implying that the emission should 
have a spectral index of 3.75.  However, numerous recent observational 
studies of protostars have found $\alpha < 3.75$ over a variety of spatial 
scales \citep[e.g.,][]{kwon2009:beta,melis2011:l1527,shirley2011:mustang,scaife2012:disks,chiang2012:l1157,tobin2013:l1527}.  
In an extreme example, \citet{tobin2013:l1527} measured $\alpha = 2$ between 
3.4 and 0.87 mm over baselines ranging from 10 -- 500 \kl, implying a flat 
dust opacity law ($\beta = 0$).

\begin{figure}
\resizebox{\hsize}{!}{\includegraphics{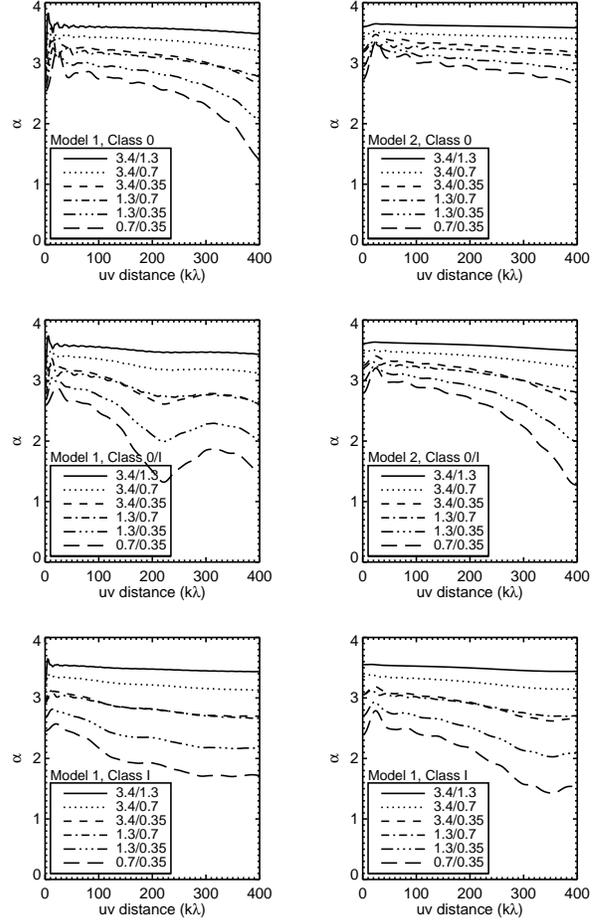}}
\caption{Spectral index, $\alpha$, calculated from each wavelength pair for 
each timestep of each model, as a function of baseline distance.  Calculations 
over different wavelength ranges are shown, with the legend in each panel 
showing the two wavelengths used to calculate $\alpha$ for each line.  A 
source inclination of 45\degree\ was assumed for these calculations.}
\label{fig_spectralindex}
\end{figure}

There are two ways to lower the spectral index: partially optically thick 
emission and grain growth.  
It is well-established that dust grains can quickly grow to 
millimeter sizes or larger in Class II discs 
(older discs than those considered in this study), 
and that such growth can change the opacity law by decreasing $\beta$ 
\citep[see][for a recent review]{testi2014:ppvi}.  Whether or not 
grains have grown sufficiently large to change the opacity law in protostellar 
discs remains an open question.
Since we have already shown that the discs 
in these models are partially optically thick, especially at shorter 
wavelengths, we first evaluate whether optical depth alone can explain these 
recent results of $\alpha < 3.75$ in observations of protostars.  
Figure \ref{fig_spectralindex} plots the spectral index, $\alpha$, calculated 
from each wavelength pair, for each timestep of each model, as a function of 
baseline distance.  Since grain growth is not an option in models with a fixed 
dust opacity law, and $\beta = 1.75$ over all submillimeter and millimeter 
wavelengths for OH5 dust opacities, 
any deviations from $\alpha = 3.75$ are due to 
optical depth effects.  We find that $\alpha$ is lower when calculated from 
shorter wavelengths that are more optically thick, and is less than 3.75 
even when calculated only from millimeter wavelengths.  Similar results are 
found when three or four wavelengths are used to calculate $\alpha$, as long 
as the first and last wavelengths remain the same.  While these results do 
suggest that optical depth lowers the observed $\alpha$, the 
\citet{tobin2013:l1527} results are most comparable to our results for $\alpha$ 
calculated from 3.4 and 0.7 mm, and since we find $\alpha \sim 3-3.5$ at these 
wavelengths, optical depth effects alone are unable to explain their results.  
Furthermore, while the source studied by 
\citet{tobin2013:l1527} is at a nearly edge-on 
inclination and our results are presented for an inclination of 45\degree, 
we find qualitatively similar results even for an inclination as high as 
85\degree\ in that the observed $\alpha$ is generally larger than two.

If grains have grown to larger sizes in discs, as suggested by combining our 
results with the observational studies cited above 
\citep[see also the discussion in][]{tobin2013:l1527}, the opacity law will 
flatten.  This will generally lead to smaller dust opacities at submillimeter 
wavelengths and larger dust opacities at millimeter wavelengths.  To give a 
specific example, the dust opacity law of \citet{ricci2010:disks}, 
which includes grains that have grown up to 1 mm, 
predicts $\beta=0.25$ and $\kappa = 2.5$ cm$^2$ g$^{-1}$ at 850 \um.  This 
opacity law is within a factor of 1.3 of that predicted by the OH5 dust 
opacity law at 850 \um\ (1.89 cm$^2$ g$^{-1}$), but leads to opacities that 
are a factor of 3 smaller at 350 \um\ and factors of 2.5 and 10.5 larger at 
1.3 and 3.4 mm, respectively.  If the \citet{ricci2010:disks} opacity law 
represented the true opacity law in the disc, but the OH5 dust opacities were 
erroneously adopted to calculate disc masses, 
the resulting masses would underestimate the true disc masses at the 
shortest wavelengths but overestimate them at the longest wavelengths.  
Future work must concentrate on determining 
whether the \citet{tobin2013:l1527} result of a very low spectral index for 
L1527 is an anomaly or the norm for protostars, and use these results to 
set detailed constraints on the opacity law for dust in protostellar discs.

\section{Discussion}\label{sec_discussion}

\subsection{Reliability of Protostellar Disc Mass Measurements}\label{sec_discussion_reliability}

\begin{figure*}
\centering
\includegraphics[width=18cm]{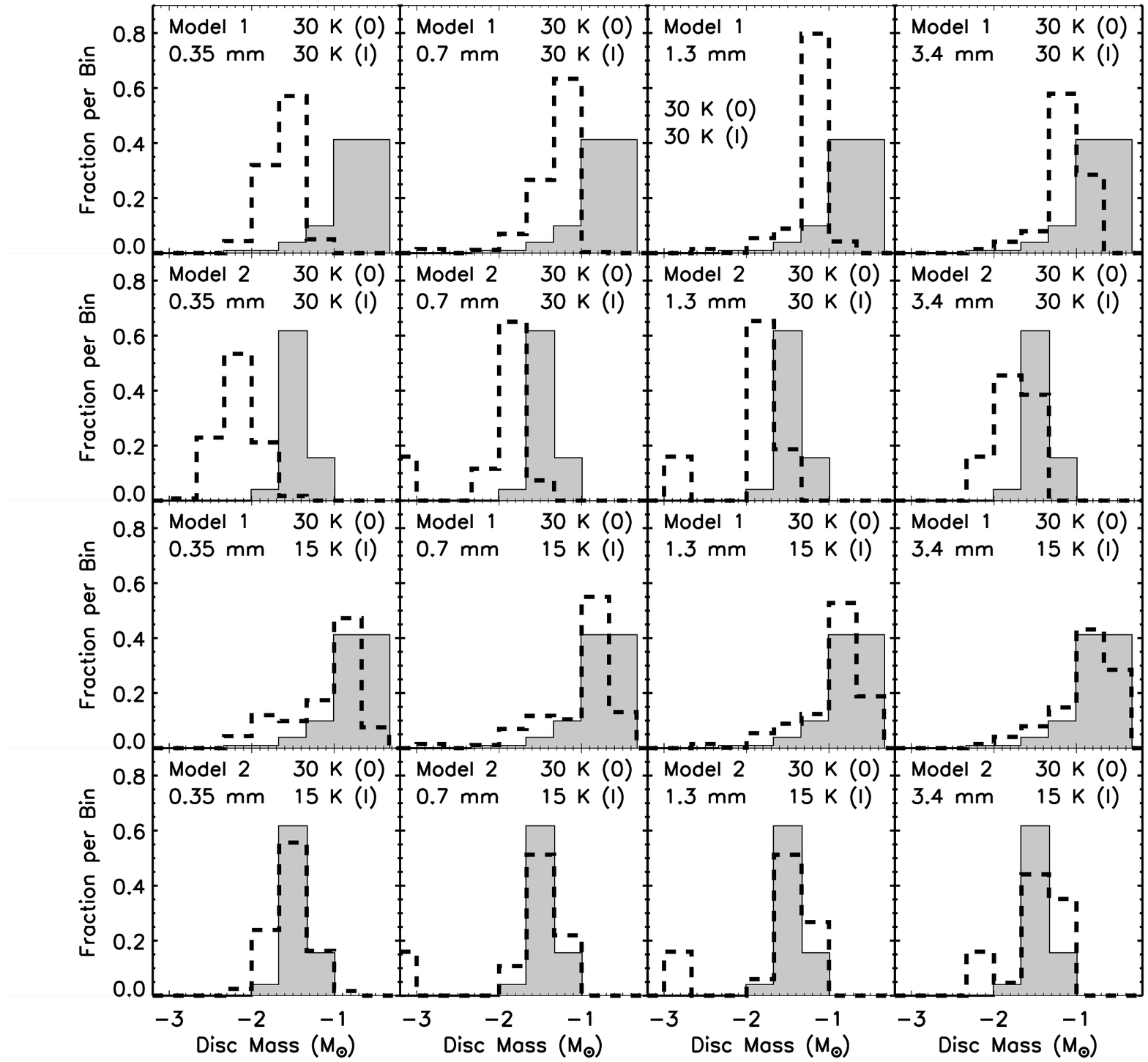}
\vspace{-1.42cm}
\caption{
Protostellar disc mass distributions for each timestep of each model.  In each 
panel the solid histograms show the intrinsic distributions taken directly 
from the simulation data, thus the solid histograms are the same in each panel 
for a given model.  The dashed histograms show the synthetic disc mass 
distributions over the full evolution of each model calculated from the 
synthetic interferometer amplitude at 50 \kl\ (see text for details).  
In the top two rows the assumed isothermal dust temperature is 30 K for 
all timesteps, whereas in the bottom two rows it is 30 K for timesteps in the 
Class 0 evolutionary stage and 15 K for timesteps in the Class I evolutionary 
stage.  Each panel is labeled with the model, wavelength, and isothermal dust 
temperatures adopted to calculate disc masses in the Class 0 and I evolutionary 
stages.
}
\label{fig_mdisc_mean}
\end{figure*}

\begin{table*}
\begin{center}
\caption{Intrinsic and Synthetic Mean Disc Masses}
\label{tab_masses}
\begin{tabular}{lcccccc}
\hline \hline
 & Intrinsic & & & & Synthetic & Mean Intrinsic/Synthetic \\
 & Mean Disc Mass & Wavelength & Class 0 T$_{\rm dust}$ & Class I T$_{\rm dust}$ & Mean Disk Mass & Disc Mass \\
Model & (\msun) & (mm) & (K) & (K) & (\msun)  & (\msun) \\
\hline
1 & 0.175 & 0.35 & 30 & 30 & 0.028 & 6.3 \\
  &       & 0.7  & 30 & 30 & 0.057 & 3.1 \\
  &       & 1.3  & 30 & 30 & 0.069 & 2.5 \\
  &       & 3.4  & 30 & 30 & 0.091 & 1.9 \\
2 & 0.032 & 0.35 & 30 & 30 & 0.009 & 3.6 \\
  &       & 0.7  & 30 & 30 & 0.014 & 2.3 \\
  &       & 1.3  & 30 & 30 & 0.017 & 1.9 \\
  &       & 3.4  & 30 & 30 & 0.021 & 1.5 \\
1 & 0.175 & 0.35 & 30 & 15 & 0.119 & 1.5 \\
  &       & 0.7  & 30 & 15 & 0.144 & 1.2 \\
  &       & 1.3  & 30 & 15 & 0.160 & 1.1 \\
  &       & 3.4  & 30 & 15 & 0.179 & 1.0 \\
2 & 0.032 & 0.35 & 30 & 15 & 0.031 & 1.0 \\
  &       & 0.7  & 30 & 15 & 0.035 & 0.9 \\
  &       & 1.3  & 30 & 15 & 0.038 & 0.8 \\
  &       & 3.4  & 30 & 15 & 0.043 & 0.7 \\
\hline
\end{tabular}
\end{center}
\end{table*}

To summarize the above sections, we find that, 
on average, the competing effects of contamination from the surrounding core 
and partially resolving the disc will tend to cancel each other 
at millimeter wavelengths but lead to underestimates at submillimeter 
wavelengths, by up to factors of 4 in the worst cases.  Optically thick 
emission, colder dust temperatures than the assumed $T_D = 30$ K, and the 
effects of inclination can lead to underestimates by up to factors of $2-3$ at 
millimeter wavelengths and up to an order of magnitude or larger at 
submillimeter wavelengths, especially 350 \um.
Finally, flatter dust opacity laws than the assumed 
OH5 opacity law, as implied by recent observations, will cause further 
disc mass underestimates at the shortest submillimeter wavelengths and 
overestimates at the longest millimeter wavelengths, with the exact magnitude 
dependent on the details of the intrinsic dust opacity law.

In order to understand the net sum of all of these effects 
(except for grain growth, which we are unable to evaluate with our fixed 
dust opacity law), Figure \ref{fig_mdisc_mean} plots various intrinsic and 
synthetic disc mass distributions for each model.  The intrinsic distributions 
are taken directly from the simulation data and represent the true disc mass 
distribution over each model run; thus they are the same 
in each panel for a given model.  The synthetic distributions are assembled by 
calculating, for each wavelength and each inclination, 
the disc mass throughout the full evolution of the model (in 2000 
yr timesteps) from the synthetic interferometer amplitude at 50 \kl, 
assuming optically thin, isothermal emission according to Equation 
\ref{eq_dustmass}.  The contribution of each 
inclination to the final distribution for each wavelength of each model is 
weighted by solid angle (see \citet{dunham2010:evolmodels} for details).  
In the top two rows the assumed isothermal dust temperature is 30 K for 
all timesteps, whereas in the bottom two rows it is 30 K for timesteps in the 
Class 0 evolutionary stage and 15 K for timesteps in the Class I evolutionary 
stage.  The mean of each distribution is listed in Table \ref{tab_masses}.

When the isothermal $T_D$ is held fixed at 30 K independent of evolutionary 
stage, the calculated disc mass distributions are clearly lower limits to 
the intrinsic distributions, with mean values lower by factors of 
$1.5-6$ depending on wavelength and model.  However, when two values of 
$T_D$ are adopted, 30 K for Class 0 sources and 15 K for Class I sources, 
the calculated distributions are much more representative of the intrinsic 
distributions (mean values within 50\%).

\subsection{Existence of Fragmenting Protostellar Discs}\label{sec_discussion_fragmenting}

\begin{figure*}
\centering
\includegraphics[width=16cm]{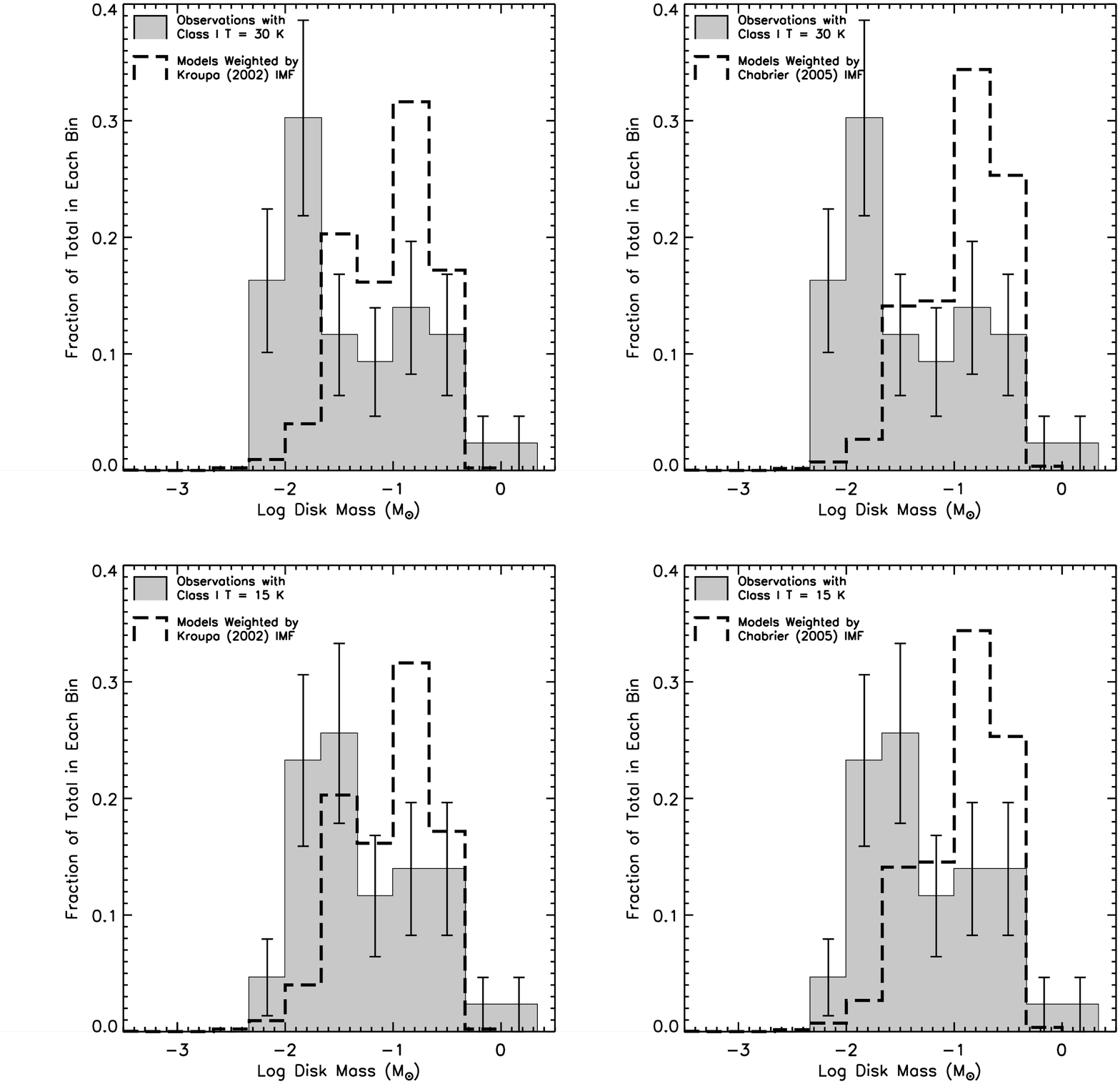}
\caption{Distribution of protostellar disc masses.  In each panel, the shaded 
histogram shows the combined distribution of observed disc masses 
from \citet{jorgensen2009:prosac}, \citet{enoch2011:disks}, and 
\citet{eisner2012:disks}, with the error bars showing the statistical 
($\sqrt{N}$) uncertainties, and the dashed histogram shows the distribution of 
protostellar disc masses predicted by the \citet{vorobyov2010:bursts} 
simulations.  The IMF used to calculate the distributions from the simulations 
are those of \citet[][left panels]{kroupa2002:imf} and 
\citet[][right panels]{chabrier2005:imf} (see text for details).
The top two panels show the observed disc mass distribution calculated assuming 
a fixed isothermal dust temperature of $T_D = 30$ K independent of evolutionary 
stage.  The bottom two panels show the observed disc mass distribution 
calculated assuming $T_D = 30$ K for Class 0 sources and $T_D = 15$ K for 
Class I sources.
}
\label{fig_mdisc_histogram}
\end{figure*}

Given the results presented in this paper, what constraints on protostellar 
disc masses can be set by existing surveys?  Figure \ref{fig_mdisc_histogram} 
plots the combined distribution of protostellar disc masses from 
\citet{jorgensen2009:prosac}, \citet{enoch2011:disks}, and 
\citet{eisner2012:disks}, all of which are based on observations at 
1.1 or 1.3 mm.  
In the top two panels, disc masses were taken directly from the first two 
surveys, who used the method evaluated here with a fixed $T_D$ of 30 K, and 
were calculated from the \citet{eisner2012:disks} survey based on their 
reported visibility amplitudes for each source, again assuming a fixed 
$T_D$ of 30 K.  We note that these calculated disc masses generally agree 
to within a factor of two with masses derived by \citet{eisner2012:disks} 
based on radiative transfer modeling of each source, which we take to be 
excellent agreement considering they fit to a relatively coarsely sampled 
model grid.  In the bottom two panels, the disc masses from all three 
surveys are recalculated assuming $T_D = 30$ K for Class 0 sources and 
$T_D = 15$ K for Class I sources.  Altogether, 43 disc mass 
measurements are plotted; an additional four non-detections in the combined 
sample are not shown in Figure \ref{fig_mdisc_histogram}.  
The observed distribution of protostellar disc masses spans more than two 
orders of magnitude, from less than 10$^{-2}$ \msun\ to greater than 1 \msun, 
with a peak slightly above 10$^{-2}$ \msun\ (although, as demonstrated 
by the error bars in Figure \ref{fig_mdisc_histogram}, the shape of this 
distribution is still quite uncertain due to small-number statistics). 
 
Figure \ref{fig_mdisc_histogram} also plots the distribution of protostellar 
disc masses predicted by the \citet{vorobyov2010:bursts} simulations, which 
often form discs that are sufficiently massive to fragment and drive 
short-lived accretion bursts from the disc onto the protostar.  
This distribution is calculated as the fraction of total time spent at each 
disc mass for an ensemble of 23 simulation runs spanning a large range of 
initial core masses and angular momenta (see Table 1 of 
\citet{dunham2012:evolmodels} for a full list).  The contribution of each 
simulation to the total distribution is weighted by either the 
\citet[][left]{kroupa2002:imf} or \citet[][right]{chabrier2005:imf} stellar 
initial mass function (IMF) based on the final mass of the star formed in each 
simulation (see \citet{dunham2012:evolmodels} for details of this weighting).

Comparing the two distributions in the top panels reveals that the relatively 
massive discs predicted by the \citet{vorobyov2010:bursts} non-magnetic 
hydrodynamical simulations are observed in approximately 50\% of all 
protostars.  However, since the observed disc mass distribution in these 
panels was assembled a fixed $T_D = 30$ K, our results from \S 
\ref{sec_discussion_reliability} imply that it is a lower limit to the 
intrinsic protostellar disc mass distribution.  Comparing the two 
distributions in the bottom panels, where $T_D = 15$ K for Class I sources 
and thus the observed distribution should closely match the intrinsic 
distribution, the discs predicted by the \citet{vorobyov2010:bursts} 
simulations are observed in approximately 70\% of all protostars.  In both the 
top and bottom panels somewhat better agreement is found when 
the simulations are weighted by the \citet{kroupa2002:imf} IMF, which gives 
increased weight to simulations that form lower-mass stars.  

Based on these comparisons, and given that the \citet{vorobyov2010:bursts} 
simulations are prone to form fragmenting discs, particularly when the disk 
mass exceeds $\sim$0.07 \msun\ \citep{Vorob2013}, we conclude 
that at least some protostellar discs are likely sufficiently massive to 
fragment.  The fact that the observed and simulation disc mass distributions 
do not show perfect overlap (in particular, the observed distribution extends 
to lower masses) may be explained by magnetic braking acting to 
remove angular momentum and suppress disc formation in some cases.  
Additionally, we caution that the surveys published to date 
are generally biased toward massive and luminous protostars that are thus 
bright at (sub)millimeter wavelengths, and may not be representative of all 
embedded protostars.  Larger, more representative surveys are needed to test 
these findings.  We are currently working on assembling such samples, on 
including the effects of magnetic braking into the numerical models, and on 
searching for other observational signatures of fragmenting discs beyond 
their total masses, such as those predicted by \citet{vorobyov2013:fragments}, 
including direct detection of spiral structure and fragments and characteristic 
SED features.

\subsection{Recommendations for Future Studies}\label{sec_discussion_recommendations}

Given the results presented in this paper, we make the following two 
recommendations for future studies that seek to use interferometer continuum 
observations of protostars to measure the protostellar disc mass distribution 
for larger samples than those considered to date:

\begin{enumerate}
\item Observe at a wavelength of approximately 1 mm.  Shorter wavelengths will 
be more optically thick and more prone to errors due to incorrect temperature 
assumptions (since they are farther outside of the Rayleigh-Jeans limit), and 
longer wavelengths will be more prone to disc mass overestimates due to 
larger opacities caused by grain growth.
\item Adopt two different isothermal dust temperatures depending on the 
evolutionary stage of each protostar: 30 K for Class 0 protostars and 15 K 
for Class I protostars.
\end{enumerate}

\section{Limitations and Caveats}\label{sec_limit}

\subsection{Physical Structure}\label{sec_limit_structure}

While both the hydrodynamical simulations and radiative transfer calculations 
are performed using two-dimensional codes, they use different coordinate 
systems.  In the hydrodynamical simulations, the thin-disc 
approximation is used to calculate the evolution of the gas surface density and 
velocity profiles as a function of cylindrical radius, $s$, and azimuthal 
angle, $\phi$, 
whereas the radiative transfer calculations assume axisymmetry to calculate the 
dust temperature distribution as a function of radius, $r$, and zenith angle, 
$\theta$, in 
spherical coordinates.  As a result, the radiative transfer calculations 
are able to match the exact mass, radius, and azimuthally-averaged surface 
density profile of the disc but do not recover its axial structure, 
in particular spiral density waves and fragments.

As described in \S \ref{sec_radtrans}, we match the disc surface density 
profiles 
in the simulations and radiative transfer calculations by fitting a power-law 
to the azimuthally-averaged surface density profile predicted by the 
simulations.  However, inspection of the simulations reveal that the profiles 
are generally better described by a broken power-law, with a much steeper 
index at large radii, than by a single power-law over the full extent of the 
disc.  We included 
such a broken power-law in the radiative transfer calculations to test its 
effects but found that it had no significant effects on our results, since 
it changes slightly the distribution of mass within the disc but not the total 
mass recovered by the simulated observations.

However, we do acknowledge that there are limitations related to our inability 
to include the full, non-axisymmetric structure of the disc in our radiative 
transfer calculations.  One such limitation is the effect of disc fragments.  
In the most extreme cases, the discs in the \citet{vorobyov2010:bursts} 
simulations form fragments that contain up to 50\% of the total mass of the 
disc.  Such fragments will be extremely optically thick and thus essentially 
invisible to (sub)millimeter interferometric observations.  
Thus, in addition to the effects discussed in \S \ref{sec_results}, 
hidden fragments will cause disc masses calculated from interferometer 
visibility amplitudes measured at 50 \kl\ to be further underestimated.  
The full magnitude of these underestimates can only be evaluated once the 
simulations are coupled to three-dimensional radiative transfer calculations 
that capture the full disc structure; we are currently working on such models 
and will present the results in a forthcoming paper.

Another limitation is the effect that grain growth will have on the assumption 
that the gas and dust are well-coupled, which we adopt when using the gas 
surface density from the simulations to set the dust surface density profile 
in the radiative transfer calculations (as described in \S 
\ref{sec_radtrans}).  If dust grains in protostellar discs have already grown 
to millimeter sizes, the dust surface density profile can deviate significantly 
from the gas due to preferential concentration of these large grains within 
density enhancements \citep[such as fragments, clumps, and spiral arms;][]{rice2004:graingrowth,boley2010:graingrowth,testi2014:ppvi}.  
Thus we again emphasize the critical need for future work devoted to 
understanding the dust properties in protostellar discs to go along with recent 
advances in our understanding of dust properties in older, more evolved 
(i.e., Class II) discs \cite[see][for a recent review]{testi2014:ppvi}.  Such 
work requires high-angular resolution, high-sensitivity, multi-wavelength 
observations of embedded protostellar discs and should thus be viable in the 
coming years with ALMA and the upgraded Very Large Array (VLA).

\subsection{Confirming Keplerian Rotation}\label{sec_limit_keplerian}

Emission on baselines of 50 \kl\ corresponds to spatial scales of approximately 
$500-2500$ AU for sources at distances ranging from $100-500$ pc, thus we 
must emphasize that such observations do not spatially resolve the discs.  
As discussed in \S \ref{sec_results_resolve}, this is actually a strength as 
it allows this method to probe the total emission from the disc, but at the 
cost of assuming that the unresolved component does in fact arise from a 
rotationally supported, Keplerian disc.  To test whether emission on 50 \kl\ 
baselines truly does imply the presence of a disc-like structure, we generated 
synthetic interferometric observations of an alternative version of model 1 
with the disc removed and the core inner radius set to 1 AU rather than the 
outer radius of the disc, which is generally 100 AU or larger (see 
Figure \ref{fig_models}).  The resulting disc masses calculated from the 
amplitudes at 50 \kl\ for this model ranged from 5 $\times 10^{-3}$ \msun\ 
early in the Class 0 stage to less than $10^{-3}$ \msun\ late in the Class I 
stage.  As most observations performed to date suggest discs with masses 
greater than $10^{-2}$ \msun\ (see Figure \ref{fig_mdisc_histogram}), 
even cores with such extreme inner radii 
have a negligible effect on the observed disc mass distribution derived 
assuming all emission on 50 \kl\ baselines arises from discs.  These 
results are in general agreement with those of \citet{jorgensen2009:prosac}, 
who also found that cores do not contribute significantly to the total 
emission at 50 \kl, even in cases where the inner radii are as small as 25 AU.  
They are also in agreement with earlier work that arrived at 
similar conclusions using much simpler models 
\citep[e.g.,][]{terebey1993:disks}.  Non-magnetized cores simply do not 
contain enough mass on the small spatial scales probed by 50 \kl\ baselines.

While these results demonstrate that the structures probed by interferometer 
observations at baselines of 50 \kl\ can not be explained as simply the inner, 
dense parts of cores, they still do not prove that the structures are due to 
rotationally supported, Keplerian discs.  Magnetically supported 
``pseudodiscs'' form on size scales of a few hundred to a few thousand AU 
in magnetized cores \citep[e.g.,][]{galli1993:bfields,chiang2008:pseudodisks}, 
and unresolved 
continuum observations are unable to distinguish between the two types of 
objects.  
We emphasize that the observed disc mass distribution  
assembled using the methods investigated in this study must be confirmed with 
detections of Keplerian rotation signatures in molecular line observations.  
While such observations are now possible 
\citep[e.g.,][]{jorgensen2009:prosac,tobin2012:l1527,murillo2013:vla1623a,yen2013:discs,harsono2014:discs}, they push existing facilities such as the 
Submillimeter Array (SMA), Combined Array for Research in Millimeter-wave 
Astronomy (CARMA), and Plateau de Bure Interferometer (PdBI) 
to their limits and are only possible for the brightest and most massive 
sources, especially in the Class 0 stage.  However, they should become routine 
in the near future with the technical capabilities of ALMA and the Northern 
Extended Millimeter Array (NOEMA).  Confirming Keplerian rotation and 
measuring the radii out to which such motions are detected is a high priority 
for future observations.

\section{Summary}\label{sec_summary}

In this paper we have coupled hydrodynamical simulations of the collapse of 
dense, rotating cores into protostars with radiative transfer 
evolutionary models to 
generate synthetic observations as the simulations evolve.  We have used 
this coupling to investigate the extent to which a simple method for 
measuring protostellar disc masses, namely the assumption that the 
(sub)millimeter interferometer amplitude measured at a baseline of 50 \kl\ 
arises from optically thin, 30 K thermal dust emission in the disc with 
no contamination from the surrounding core, recovers the intrinsic masses of 
the discs formed in the simulations.  We summarize our main results as follows:

\begin{itemize}
\item Overestimates due to contamination from the surrounding core and 
underestimates due to partially resolving out the disc tend to cancel each 
other at millimeter wavelengths but can lead to net underestimates by up to 
factors of 4 at submillimeter wavelengths.
\item Optically thick emission, colder dust temperatures than the assumed 
$T_D = 30$ K, and the effects of inclination can lead to underestimates by 
up to factors of 2--3 at millimeter wavelengths and up to an order of magnitude 
or larger at submillimeter wavelengths, especially 350 \um.  
\item Flatter dust opacity laws due to grain growth will lead to further 
disc mass underestimates at the shortest submillimeter wavelengths and 
overestimates by up to factors of 10 or larger at the longest millimeter 
wavelengths (3 mm or longer), with the exact magnitude dependent on the 
details of the true opacity law in the disc.
\item The optically thin portions of protostellar discs are generally cooler 
in the Class I stage than the Class 0 stage since Class I discs are typically 
larger and more optically thick, and thus more shielded. Protostellar disc mass 
distributions assembled from observations where the dust temperature is fixed 
at 30 K independent of evolutionary stage will be lower limits to the 
intrinsic distributions, with mean values lower by factors of $1.5-6$ 
depending on wavelength and intrinsic disc properties.  
If temperatures of 30 K for Class 0 sources and 15 K for Class I sources are 
adopted, the observed distributions will be more representative of the 
intrinsic distributions, with mean values within 50\%.
\item Comparing the observed disc mass distribution from three different 
surveys at 1 mm to that assembled from 
the \citet{vorobyov2010:bursts} hydrodynamical simulations suggests that 
the disc masses predicted by the simulations are observed in 
approximatley 50\% -- 70\% of all protostars.  
These results suggest that at least some protostellar discs are likely 
sufficiently massive to fragment, although larger samples are needed to 
further quantify this result.
\item We make the following two recommendations for future studies that seek 
to use interferometer continuum observations of protostars to measure the 
protostellar disc mass distribution:
\begin{enumerate}
\item Observe at a wavelength of approximately 1 mm.
\item Adopt a variable dust temperature of 30 K for Class 0 sources and 15 K 
for Class I sources.
\end{enumerate}
\end{itemize}

These results must be confirmed with larger, less biased observational 
samples.  Additionally, this method is not capable of distinguishing between 
rotataionlly supported, Keplerian discs and magnetically supported 
``pseudodiscs'', thus future molecular line observations are critically needed 
to confirm Keplerian rotation in protostars found to have discs with the 
method investigated here.  However, the current evidence does support the 
hypothesis that disc fragmentation may play a significant role in the early 
evolution of protostellar systems.

We thank the anonymous referee for comments and suggestions 
that have improved the quality of this work.  
This research has made use of NASA's Astrophysics Data System (ADS) 
Abstract Service and the IDL Astronomy Library hosted by the NASA Goddard Space 
Flight Center.  MMD acknowledges support from the Submillimeter Array through 
an SMA postdoctoral fellow.  The simulations were performed
on the Shared Hierarchical Academic Research Computing Network
(SHARCNET), on the Atlantic Computational Excellence Network (ACEnet),
and on the Vienna Scientific Cluster (VSC-2).  
This project was partly supported by the Russian Ministry of Education and 
Science Grant (state assignment) 3.961.2014/K.  
HGA acknowledges support from the NSF through grant AST-0845619.

\bibliographystyle{mn2e.bst}
\bibliography{dunham_citations}

\end{document}